\documentclass{iopart}
\usepackage{epsfig}
\usepackage{latexsym}

\begin{document}

\newcommand{\hide}[1]{}
\newcommand{\tbox}[1]{\mbox{\tiny #1}}
\newcommand{\mbf}[1]{{\mathbf #1}}
\newcommand{\half}{\mbox{\small $\frac{1}{2}$}}
\newcommand{\sinc}{\mbox{sinc}}
\newcommand{\const}{\mbox{const}}
\newcommand{\trc}{\mbox{trace}}
\newcommand{\ointt}{\int\!\!\!\!\int\!\!\!\!\!\circ\ }
\newcommand{\eexp}{\mbox{e}^}
\newcommand{\bra}{\left\langle}
\newcommand{\ket}{\right\rangle}


\title
[Survival probability and LDOS for 1D systems]
{Survival probability and local density of states for one-dimensional Hamiltonian systems}

\author{Ji\v{r}\'{\i} Van\'{\i}\v{c}ek$^{1,2}$ and Doron Cohen$^3$}

\date{March 2003}

\address{
$^1$ \mbox{Department of Physics, Harvard University, Cambridge, MA 02138} \\
$^2$ \mbox{Mathematical Sciences Research Institute, Berkeley, CA 94720} \\
$^3$ \mbox{Department of Physics, Ben-Gurion University, Beer-Sheva 84105, Israel}
}


\begin{abstract}
For chaotic systems there is a theory for the decay of
the survival probability, and for the parametric dependence of
the local density of states. This theory leads to the distinction
between ``perturbative" and ``non-perturbative" regimes,
and to the observation that semiclassical tools are useful
in the latter case. We discuss what is ``left" from this
theory in the case of one-dimensional systems.
We demonstrate that the remarkably accurate {\em uniform} semiclassical
approximation captures the physics of {\em all} the different regimes,
though it cannot take into account the effect of strong localization.
\end{abstract}

\section{\label{introduction}Introduction}

The quantum mechanical state of a particle, or of a system,
is represented by the probability matrix $\rho$.
This object corresponds to a classical distribution $\rho^{cl}(Q,P)$
in phase space, where $Q$ and $P$ are the canonical coordinates
of the system. One way to represent the probability matrix is
by using the Wigner function $\rho(Q,P)$.
Many calculations in quantum mechanics reduce eventually
to calculation of a trace over a pair of probability matrices.
This includes in particular calculations of the local density of states (LDOS) \cite{wls},
and calculations of the survival probability ${\cal P}(t)$ \cite{heller}.
Strongly related are calculations of Franck-Condon factors for
non-adiabatic transitions between Born-Oppenheimer surfaces \cite{bilha}.
Lately \cite{fidelity,jacq,fdl,fdl2} there is a special interest in calculation of ``fidelity"
(also known as ``Loschmidt echo") in the context of quantum computation.
This ``fidelity" is in fact a synonym for a ``survival probability" ${\cal P}(t)$
which is calculated for a particular dynamical scenario (namely, the forward evolution
is followed by a reversed evolution with a perturbed Hamiltonian).

The calculation of a trace over a pair of probability matrices
has a clear classical limit. To be specific let us consider
the survival probability, which is defined as
\begin{eqnarray} \label{e1}
{\cal P}(t)  = | \langle \psi_0 | \psi_t \rangle |^2
= \trc (\rho_t \rho_0).
\end{eqnarray}
where $\rho_t$  and $\psi_t$ are the probability matrix
and the associated wavefunction of an evolving
quantum mechanical pure state.
In the Wigner representation the trace operation
means $dQdP/(2\pi\hbar)^d$ integral over phase space,
where $d$ is the number of freedoms.
It should be emphasized that the Wigner representation
is quantum-mechanically exact. The classical limit/approximation
is obtained by treating $\rho(P,Q)$ as a classical distribution,
whose evolution is governed by classical  equations of motion.

To take the classical limit as a leading order
approximation for ${\cal P}(t)$ is one possibility.
Another possibility is to use perturbation theory,
in particular ``Fermi golden rule" (FGR).
It should be obvious that the results that are
obtained by using these methods are typically
very different from each other.

A major objective of recent studies \cite{ovrv}
is to understand the limitations of perturbation theory on the one hand,
and to explore the capabilities of the semiclassical tools, on the other.
A specific question is whether the decay of ${\cal P}(t)$ is
of perturbative nature (e.g. FGR type
\footnote{
In section~5 we explain that perturbation theory can
lead to either no decay, or Gaussian decay, or Wigner type decay.
The latter can be regarded as the outcome of FGR transitions.
}),
or else whether it is of classical nature ("semiclassical" type).
The main approach towards this question is to allow
the specification of a control parameter that represents
the ``strength" of a perturbation.
Depending on the value of of this control parameter
one may have either perturbative or semiclassical behavior.
The first publications  \cite{crs,wls}  that have taken this approach,
led to the distinction between ``perturbative" and ``non-perturbative"
regimes, and to the realization that the ``semiclassical" behavior is
contained in the"non-perturbative" regime.
Later \cite{jacq}  the idea was adopted into the context of ``fidelity" studies.

The above mentioned studies have mainly concentrated on
quantized {\em chaotic} systems. The chaos assumption allows
simplification of certain calculations. In particular one
can invoke the random matrix theory (RMT) conjecture,
in order to obtain some ``generic" results.
The natural question that arises is whether some
of the general theory regarding the LDOS and ${\cal P}(t)$,
applies also to the world of one dimensional ($d=1$) systems.

A central observation in the theory of quantized chaotic
systems is the existence of two distinct energy scales.
One is the mean level spacing ($\Delta\propto\hbar^d$),
while the other is the bandwidth ($\Delta_b \propto \hbar$).
The latter is semiclassically related to the correlation time
of the classical motion. The dimensionless bandwidth
is defined as $b=\Delta_b/\Delta$. The classical limit
($\hbar\rightarrow0$) of quantized chaotic system ($d>1$)
is characterized by  $\Delta\ll\Delta_b$ and hence  $b\gg1$.
But in one dimensional systems ($d=1$) we do
not necessarily have this separation of energy scales.
For some typical perturbations $b={\cal O}(1)$.
We shall explain that in such case there is no FGR regime
in the theory. A necessary, but not sufficient condition
for having FGR regime in one-dimensional systems is to have $b\gg 1$.
This means that ``small features" should characterize
the phase space manifolds that support the perturbed
(or evolving) quantum mechanical states.

The analysis of one dimensional  ($d=1$) systems is highly
non-universal, but typically allows analytical calculations
that go well beyond the leading semiclassical approximation.
In particular we demonstrate  the capabilities and the limitations
of the uniform approximation \cite{jiri,jiri2}.

The outline of this paper is as follows:
In Section~\ref{models} we define some simple prototype models.
In Section~\ref{definitions} we define the main objects of the
studies which are the LDOS and the survival probability~${\cal P}(t)$.
In Section~\ref{qcc} we discuss the issue of quantum-classical
correspondence (QCC) and make the distinction
between ``semiclassical" and ``perturbative" approximations.
In Sections~\ref{regimes1} and~\ref{regimes2} 
we discuss the notion of ``regimes"
in the context of LDOS studies. In Sections~\ref{large_bandwidth} 
and~\ref{uniform_regimes} we discuss the existence of LDOS regimes
in 1D within the framework of a uniform approximation.
In Section~\ref{strong_localization} we discuss the implication of the strong
localization effect. In Section~\ref{survival} we extend the general
consideration into the context of survival probability studies.
Conclusion and final remarks are summarized in Section~\ref{conclusions}.
The appendices are an integral part of the Paper.
They contain details of derivations and have not been
integrated into the main text in order to simplify the
presentation of the physical picture.

\section{\label{models}Definition of the models}

In this paper we are going to consider one-dimensional model systems.
The simplest is the deformable harmonic oscillator:
\begin{eqnarray} \label{e2}
{\cal H}(Q, P; x) = \half (x P)^2 + \half (Q / x)^2.
\end{eqnarray}
where $(Q,P)$ are the canonical coordinates,
and $x$ is a constant parameter. The energy surfaces
${\cal H}(Q, P; x) =E$ are ellipses in phase space.
We shall assume that $E\gg1$.
Without loss of generality we shall regard $x=x_0=1$
as the ``unperturbed" value of the parameter $x$.
Later we shall assume that $\delta x \equiv (x-x_0)$
is small in a {\em classical} sense. Namely $\delta x \ll 1$.
Then it is possible to write the Hamiltonian as
${\cal H}={\cal H}_0+\delta {\cal H}$, where
\begin{eqnarray} \label{e3}
{\cal H}_0  \  &= \ {\cal H}(Q,P;x_0{=}1) \ =  \ \frac{1}{2} (P^2+Q^2),
\\  \label{e4}
\delta {\cal H} \  &= \
\delta x \left.\frac{\partial {\cal H}}{\partial x}\right|_{x=1}  \ = \ \delta x  \ (P^2-Q^2)
\end{eqnarray}
Note that the perturbation $\delta x$ is allowed to be
large in a quantum mechanical sense: many levels
can be ``mixed" by the perturbation $\delta {\cal H} $.

A more complicated case is to consider a particle in a ring
($=$ one dimensional box with periodic boundary conditions).
The Hamiltonian is of the general form
\begin{eqnarray} \label{e5}
{\cal H}(Q, P; x) = \frac{1}{2{\mathsf m}} P^2 + x \ V(Q)
\end{eqnarray}
Without loss of generality we can take ${\mathsf m}=1$
as the mass of the particle, and $L=2\pi$
as the perimeter (length) of the ring.
The parameter $x$ controls the ``height" of the
potential landscape. Naturally we shall regard $x_0=0$
as the unperturbed value of $x$.
We shall consider the case of a single ``bump" where
\begin{eqnarray} \label{e6}
V(Q)    =  V_0\exp \left[-\left(Q-Q_0 \right)^2 / 2 \ell^2 \right]
\end{eqnarray}
It is implicitly assumed that $\ell \ll L$.
The Fourier components of this potential are
$|\tilde{V}(k)| =  V_0\ell \exp\left[-(k\ell)^2 /2 \right]$.
Hence the non-vanishing ($|k\ell| < 1$) Fourier components
satisfy $|\tilde{V}(k)| \approx V_0\ell$.
If we have many bumps, then we have a ``disordered" potential
\begin{eqnarray} \label{e7}
V(Q) \ = \  \sum_{\alpha} (\pm\mbox{random}) V_0 \exp \left[-\left(Q-Q_{\alpha} \right)^2 / 2 \ell^2 \right]
\end{eqnarray}
Here we implicitly assume that the bumps are non-overlapping
and randomly distributed along the ring, such that the correlation
function $\langle V(Q+r)V(Q) \rangle$ is characterized by
the correlation length $\ell$. Consequently the non-vanishing
($|k\ell| < 1$) Fourier components of the potential
satisfy $|\tilde{V}(k)| \approx V_0\ell \times \sqrt{L/\ell}$.
Note that the phases of the Fourier components
look ``random", unlike the case of a single bump.

All models that we have introduced are
of the generic form ${\cal H}= {\cal H}_0 +\delta {\cal H}$.
The representation in the basis that is determined by the
unperturbed Hamiltonian is
\begin{eqnarray} \label{e8}
{\cal H} \mapsto \mbf{E} + \delta x \mbf{B}
\end{eqnarray}
where $\mbf{E}=\{E_n\}$ is a diagonal matrix
consisting of the eigenenergies of the unperturbed Hamiltonian.
With scaled parameters, such that $\hbar=1$,
the mean level spacing of the eigenenergies
is $\Delta=1$ for Hamiltonian (\ref{e2})
and $\Delta = \sqrt{2E}$  for Hamiltonian (\ref{e5}).
The matrix $\mbf{B}$ corresponds to the perturbation.
It is a banded matrix. The bandwidth is $b$. It is defined such that
$2b$ is the number of the levels which are coupled by the
perturbation. For the Hamiltonian (\ref{e2}) we have $b=1$,
because only neighboring levels of the de-symmetrized Hamiltonian ($|n-m|=2$)
are coupled by the perturbation $\delta {\cal H}$.
For the Hamiltonian (\ref{e5}) the bandwidth is $b=L/\ell$.
See Section~\ref{large_bandwidth} for details. Thus for the latter model we may have $b\gg1$.

\section{\label{definitions}Definitions of $P(n|m)$  and ${\cal P}(t)$ }

Let $|n(x)\rangle$  and  $E_n(x)$  be the eigenstates and the corresponding
eigenvalues of the Hamiltonian ${\cal H}(Q,P;x)$.
We define the ``parametric" kernel
\begin{eqnarray}
P(n|m) \ = \  | \langle n(x) | m(x_0) \rangle |^2 =  \trc (\rho_n \rho_m).
\label{ldos}
\end{eqnarray}
Note that for $\delta x=0$ we have $P(n|m)=\delta_{nm}$.
Given a reference state $|m(x_0)\rangle$ this kernel
can be regarded as a probability distribution with respect to $n$.
The LDOS is just a scaled version of this kernel, namely
\begin{eqnarray} \label{e10}
P(\omega)  \ = \ \sum_n P(n|m) \ 2 \pi \delta\Big( \omega-\left[E_n(x){-}E_m(x_0)\right] \Big)
\end{eqnarray}
One important measure that characterizes the LDOS
is its variance:
\begin{eqnarray} \label{e11}
\delta E^2 \ = \ \sum_n P(n|m) \  (E_n-E_m)^2
\end{eqnarray}
In the next section we shall explain that the variance has
a special  role in the theory of the LDOS.

The survival probability of the state $|m(x_0)\rangle$ for evolution
which is generated by the perturbed Hamiltonian ${\cal H}(Q,P;x)$
can be written as ${\cal P}(t)=|F(t)|^2$, where
\begin{eqnarray} \label{e12}
F(t) \ = \  \langle m(x_0) | \eexp{-it({\cal H}-E_m)} | m(x_0) \rangle
\end{eqnarray}
In the above definition we have taken $E=E_m$ as
the natural reference for the energy. With this definition we
see that $F(t)$ is just the Fourier transform of the LDOS.
We note that in more complicated scenarios
a simple Fourier transform relation between $F(t)$
and $P(\omega)$ does not exist \cite{fdl}.

\section{\label{qcc}Quantal-classical correspondence (QCC)}

The classical limit/approximation for the LDOS kernel $P(n|m)$ is obtained
by taking for the Wigner function  $\rho_n(Q,P)  \approx \rho^{cl}_n(Q,P)$,
where \mbox{$\rho^{cl}_n(Q,P)  \propto \delta\left[{\cal H}(Q,P;x)-E_n(x)\right] $}
is the corresponding classical microcanonical distribution.
For the deformable harmonic oscillator one obtains (\ref{appA}):
\begin{eqnarray} \label{e13}
P^{cl}(n|m) \ \approx \ \frac{1}{\pi}
\frac{1}{\sqrt{ 4( \delta x / x )^2E^2 -(E_n-E_m)^2 }}
\end{eqnarray}
The dispersion (square root of the variance) is
\begin{eqnarray} \label{e14}
\delta E^{cl}  \ = \ \sqrt{2} \left(\frac{\delta x}{x}\right) E
\end{eqnarray}

What about the quantum mechanical LDOS?
The simplest limit that can be considered
is first order perturbation theory (FOPT).
One obtains (\ref{appB}):
\begin{eqnarray} \label{e15}
P^{\tbox{prt}}(n|m) \ \approx \ \delta_{nm} \ + \
\frac{1}{4} \left(\frac{\delta x}{x}E\right)^2 \  \delta_{|n-m|,2}
\ \ \ \ \ \ \ \ \ \mbox{[FOPT]}
\end{eqnarray}
This result is very different from
the classical result. Therefore we say
that there is no ``detailed QCC". On the other
hand one easily calculates the variance.
One obtains $\delta E = \delta E^{cl}$.
We call the latter equality ``restricted QCC".

\begin{figure}
\centerline{\epsfig{figure=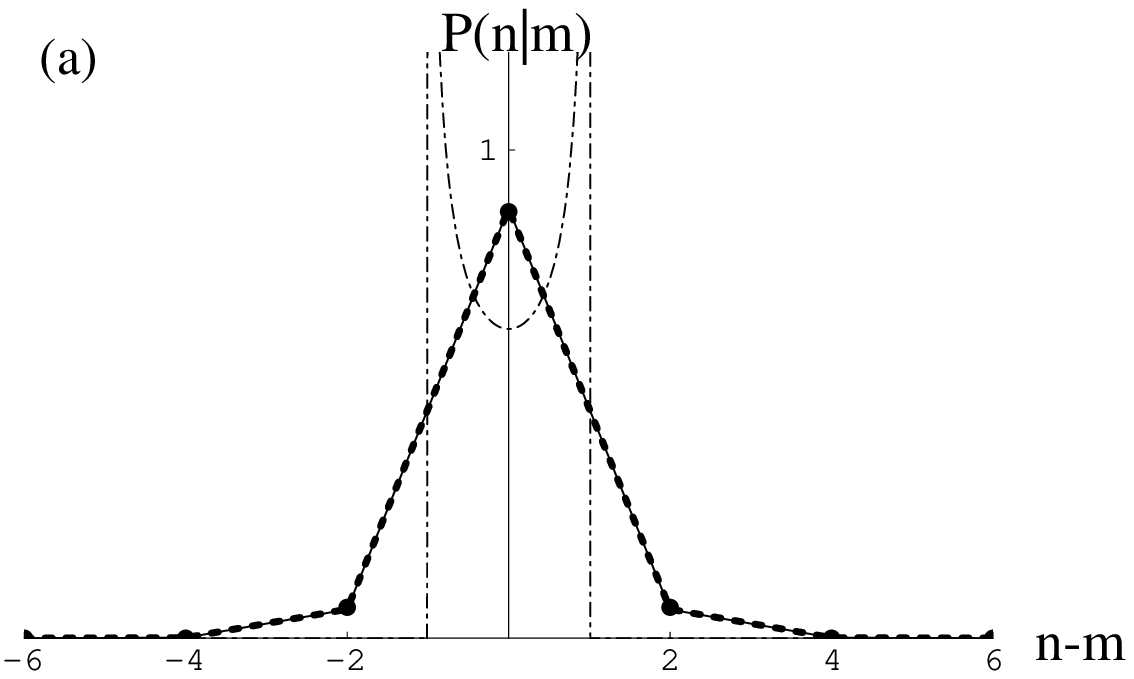,width=0.7\hsize}} \ \\
\centerline{\epsfig{figure=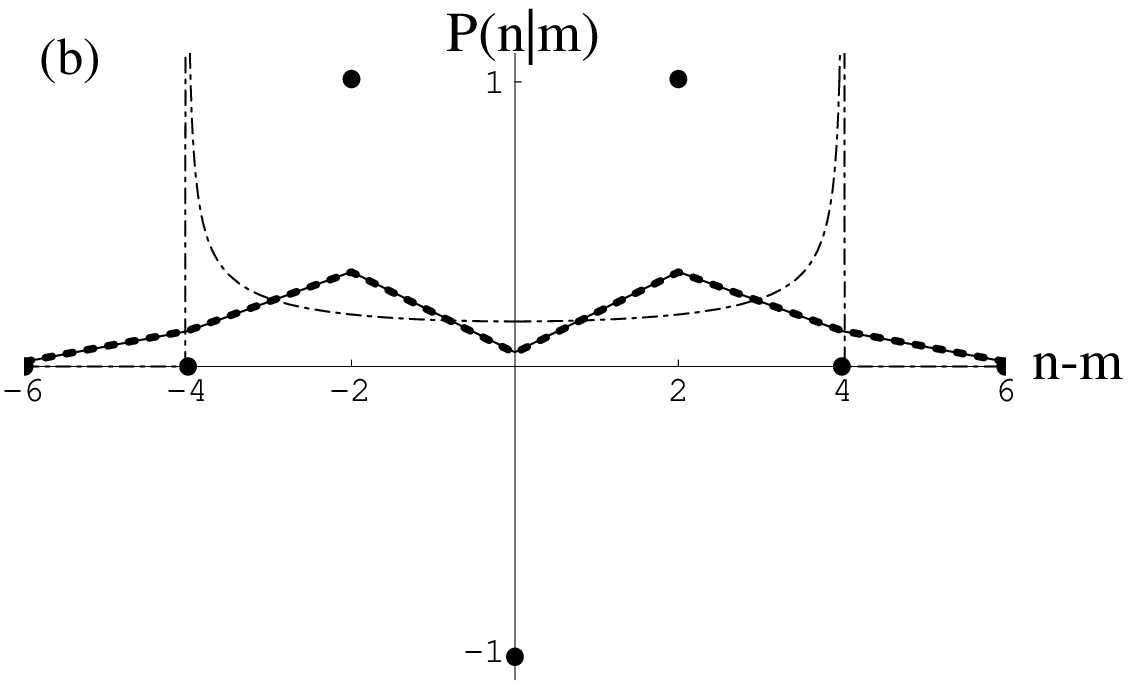,width=0.7\hsize}} \ \\
\centerline{\epsfig{figure=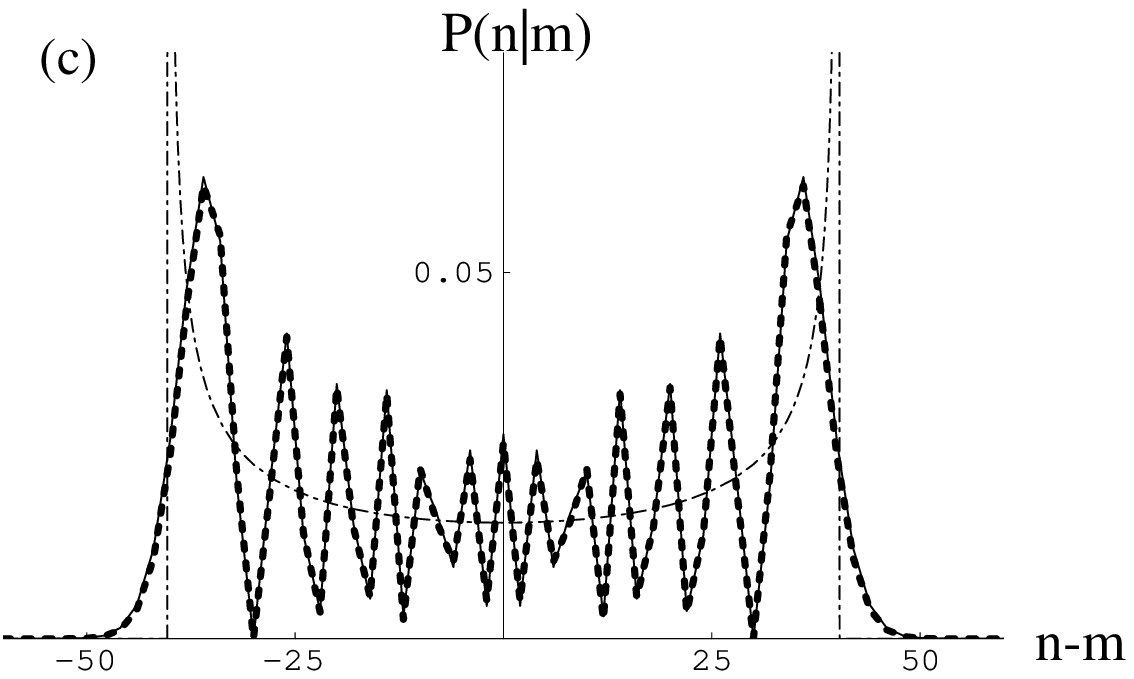,width=0.7\hsize}} \ \\

\caption{\label{fig_ldos}
The local density of states for E=100.
(a) For $(\delta x / x) E= 0.5$ (perturbative regime).
(b) For $(\delta x / x) E= 2$ (intermediate regime).
(c) For $(\delta x / x) E= 20$ (semiclassical regime).
In each panel the thick dashed line is the exact result,
the solid line is the uniform result,
the dashed-dotted line is the classical result,
and the solid circles are the perturbative result.
The classical result is divided by 2 for a reason
which is explained at the end of \ref{appA}.
}
\end{figure}

It is possible to write an exact analytical expression for
the quantum mechanical LDOS (see \ref{appA}).
But this expression is not very useful since its complexity makes
it virtually impossible to extract the simple limits in various regimes.
It is more illuminating
to obtain a uniform approximation (\ref{appC})
\begin{eqnarray} \label{e16}
P^{\rm uniform}(n|m) \ \approx \ \left[ J_{(m-n)/2} \left( \frac{\delta x}{x}E \right) \right]^2
\end{eqnarray}
In figure~\ref{fig_ldos} we demonstrate that the uniform approximation
is almost indistinguishable from the exact result for any value of~$\delta x$.
One can verify that this approximation
reduces to the FOPT result (\ref{e15})
in the parametric regime $\delta x \ll (x/E)$.
Disregarding an oscillatory component it
reduces to the semiclassical result (\ref{e13})
in the parametric regime $\delta x \gg (x/E)$.
The two regimes are separated by the quantum mechanical
parametric scale $\delta x_{\tbox{prt}} = (\hbar\omega_{osc}/E) x $.
For the convenience of the reader we have
reverted here to non-scaled  units (with our scaling
$\omega_{osc}=1$ and $\hbar=1$.)

The example above demonstrates some general
observations regarding the LDOS. On the one
hand we have restricted QCC, which means
$\delta E = \delta E^{cl}$. It can be proved that
this type of QCC holds in general \cite{lds}.
The proportionality $\delta E \propto \delta x$
is guaranteed by the classical linearization condition, 
which is a fixed assumption in this paper 
(in the present example it means $\delta x \ll x$). 
On the other hand we do not have in general detailed QCC,
which means that the approximation $P(n|m) \approx P^{cl}(n|m)$
holds only in a specific parametric regime.
In the above example we have only two parametric regimes: \\
\begin{minipage}{\hsize}
\vspace*{0.1cm}
\begin{itemize}
\setlength{\itemsep}{0cm}
\item
The ``perturbative regime"  ($\delta x \ll \delta x_{\tbox{prt}}$) \\
in which (in this example) FOPT can be used.
\item
The ``non-perturbative regime"  ($\delta x > \delta x_{\tbox{prt}}$) \\
in which (in this example) the semiclassical approximation can be used.
\end{itemize}
\vspace*{0.0cm}
\end{minipage}
Hence upon quantization there is only one parametric scale
in the theory \mbox{($\delta x_{\tbox{prt}} \propto \hbar$)}.
This parametric scale marks a border between the
perturbative and the non-perturbative regimes.

\section{\label{regimes1}Regimes in case of chaotic systems}

In case of chaotic systems the generic case is having
three regimes. The ``perturbative regime" is subdivided
into a ``FOPT regime" and
a Wigner/Lorentzian/core-tail/FGR regime.
In the latter case the various names mean the same,
and we shall use from now on the term ``Wigner regime".
The FOPT regime is defined as
the parametric regime where we can use first order
perturbation theory (FOPT):
\begin{eqnarray} \label{e17}
P^{\tbox{prt}}(n|m) \ \approx \ \delta_{nm} \ + \
\delta x^2  \frac{\ |\mbf{B}_{nm}|^2}{(E_n{-}E_m)^2}
\ \ \ \ \ \ \ \ \ \ \mbox{[FOPT]}
\end{eqnarray}
The condition for the validity of this approximation
is $\delta x \ll \delta x_c$,  where
$\delta x_c = \Delta/\sigma$.
Here $\Delta$ is the mean level spacing,
and $\sigma$ is defined as the RMS  value
of the ``in-band" off-diagonal
elements of the $\mbf{B}$ matrix.

Outside of the ``FOPT regime" we can still use
first order perturbation theory for the {\em tails} of the LDOS \cite{vrn}.
This leads to a generalized Wigner's Lorentzian
(a more meaningful name is ``core-tail structure"):
\begin{eqnarray} \label{e18}
P^{\tbox{prt}}(n|m) \ \approx \ \delta x^2
\frac{|\mbf{B}_{nm}|^2}{\left[\Gamma(\delta x)/2\right]^2 + (E_n{-}E_m)^2}
\ \ \ \ \ \ \ \ \ \ \mbox{[Wigner]}
\end{eqnarray}
The width parameter $\Gamma(\delta x)$ is determined
so as to have proper normalization \mbox{($\sum_n P(n|m)=1$)}.
For strongly chaotic systems,
for which the band profile is quite flat,
it follows that the ``core" width is
$\Gamma\approx (\delta x/\delta x_c)^2\Delta$.
The ``core-tail"  approximation makes sense
as long as we have separation of energy scales
$\Gamma(\delta x) \ll \Delta_b$,
where $\Delta_b=b\Delta$ is the bandwidth in units of energy.
This translates into the condition
$\delta x \ll \delta x_{\tbox{prt}}$ where
\begin{eqnarray} \label{e18b}
\delta x_{\tbox{prt}} \ \ = \ \ \sqrt{b} \ \delta x_c.
\end{eqnarray}

The ``core-tail"  approximation (\ref{e18})
can be regarded as the outcome of
perturbative summation of diagrams
to infinite order: The FOPT diagrams are
re-iterated, while the interference
terms are neglected. In time domain
analysis this corresponds to a {\em Markovian approximation}
for the survival probability ${\cal P}(t)$,
leading to an exponential decay.
Hence we say that there is a {\em perturbative regime}
that includes the FOPT sub-regime, and the Wigner sub-regime.
In the Wigner sub-regime we need {\em all orders}
of perturbation theory leading to FGR transitions and  Wigner decay.
On the other hand in the FOPT sub-regime there is {\em no decay}.
This can be easily deduced by Fourier transforming 
the LDOS which is associated with Eq.(\ref{e17}):
Up to negligible first order correction, the survival amplitude $F(t)$
comes out as a pure phase factor, whose absolute value squared is ${\cal P}(t)\approx1$.
It is appropriate at this point to make a connection with
recent fidelity studies \cite{fidelity}. There, it is customary to consider
the dynamics of a wavepacket that is a superposition of many eigenstates.
Consequently, in fidelity studies, there is an effective
averaging over the survival  amplitude $F(t)$,
leading (in the FOPT regime) to a slow Gaussian decay which is trivially related
to the statistics of the first order correction to the eigenenergies.

We can summarize this section by stating that in case 
of a generic quantized chaotic system we have three 
regimes as follows: \\
\begin{minipage}{\hsize}
\vspace*{0.1cm}
\begin{itemize}
\setlength{\itemsep}{0cm}
\item
The FOPT regime ($\delta x \ll \delta x_c$) \\
where Eq.(\ref{e17}) can be trusted.
\item
The Wigner (perturbative) regime ($\delta x_c \ll \delta x \ll \delta x_{\tbox{prt}}$), \\
where Eq.(\ref{e18}) can be trusted.
\item
The non-perturbative regime ($\delta x > \delta x_{\tbox{prt}}$), \\
in which (typically) the semiclassical approximation can be used.
\end{itemize}
\vspace*{0.0cm}
\end{minipage}
Hence upon quantization there are two parametric scales 
in the theory, which are $\delta x_c$ 
and \mbox{$\delta x_{\tbox{prt}} \propto \hbar$}. 
These parametric scale mark the borders between the regimes.

\section{\label{regimes2}Regimes in case of 1D systems}

By Weyl law we know that the mean level spacing
for $d$-dimensional system is $\Delta\propto\hbar^d$.
On the other hand the bandwidth $\Delta_b\propto\hbar$
is inversely proportional to the period, or to the
correlation time of the dynamics.
For chaotic systems (with $d>1$)
we generally have $b = \Delta_b/\Delta \gg1$.
But for $d=1$ systems we can have $b={\cal O}(1)$,
as in the example of equation~(\ref{e2}).
Therefore in the latter case $\delta x_{\tbox{prt}} \sim \delta x_c$,
and we do not have a ``Wigner regime". 
This means that the LDOS cannot have a Lorentzian-like structure, 
and consequently the survival probability ${\cal P}(t)=|F(t)|^2$ 
cannot have an exponential decay. 
While this latter statement strictly holds for the
specific scenario which has been defined by equation~(\ref{e12}), 
it typically holds also in more complicated  circumstances (``fidelity" studies).
Thus we conclude that the  FGR picture cannot be applicable
to the analysis of the dynamics unless $b\gg1$.

At this stage of the discussion we can say that in order
to have three regimes (FOPT, Wigner, semiclassical) in the
theory of one-dimensional systems, we have to consider
models where the perturbation is characterized by a large bandwidth ($b\gg1$).
In the next sections we would like to further discuss the following: \\
\begin{minipage}{\hsize}
\vspace*{0.2cm}
\begin{itemize}
\item
A uniform approximation can be applied in
order to address both the perturbative
and the non-perturbative regimes.
\item
We can have $b\gg1$ by considering
potentials that have small (or sharp) features.
(For example ``bump" or ``disorder").
\item
It is possible to get a Lorentzian-like LDOS
from the uniform approximation,
but $b\gg1$ is not a sufficient condition.
\item
The uniform approximation does not take
into account the effect of strong localization.
\item
Three LDOS regimes can be observed in case of disordered potential
in spite of the strong localization effect.
\end{itemize}
\vspace*{0.0cm}
\end{minipage}

\section{\label{large_bandwidth}Perturbations with large bandwidth}

For any one-dimensional  system we can find the action-angle variables,
in which the problem becomes essentially equivalent  to the
problem (\ref{e5}) of a particle in a ring.
Consider for example the deformable harmonic oscillator (\ref{e2}):
We already saw that all the essential physics of the perturbed
eigenstates can be obtained via a uniform approximation
which is based on action-angle variables description of the system.
The Hamiltonian (\ref{eC3}) of a deformable harmonic oscillator
in action-angle variables,
and the Hamiltonian (\ref{e5}) of a particle in a ring,
are similar as far as the semiclassical
treatment is concerned. Having a quadratic rather
than a linear dispersion relation is not an essential difference.

More generally, we may encounter circumstances where
the perturbation creates small phase space structures.
This is the new ingredient (compared with the deformable
harmonic oscillator) that we are going to consider in
the following sections. The problem (\ref{e5}) of a particle in a ring
constitutes a prototype example of having such small phase
space structures. The perturbation $V(Q)$ is characterized
by a different, much smaller scale ($\ell$), compared with
the size of the system ($L$). This means that we have
large bandwidth ($b\gg1$) rather than $b={\cal O}(1)$.

In the absence of a perturbation ($x=x_0=0$),
the eigenenergies of a particle in a ring are
$E_n=p_n^2/ 2 {\mathsf m}$,
where $n$ is an integer index, and $p_n = (2\pi\hbar/L) n$.
The perturbation matrix is related to the
Fourier components of the potential:
\begin{eqnarray} \label{e19_a}
\mbf{B}_{nm} \ = \ \frac{1}{L} \tilde{V}(p_n-p_m)
\end{eqnarray}
The expressions for $\tilde{V}(k)$ in case of either
the ``bump" (\ref{e6})
or the ``disorder" (\ref{e7}),
were given in Section~\ref{models}.
Substitution in (\ref{e19_a})
allows to determine the bandwidth:
\begin{eqnarray} \label{e19}
\Delta \   &=& \ 2\pi\hbar v_E/L
\\ \label{e20}
\Delta_b \ &=& \ 2\pi\hbar v_E/\ell
\\ \label{e21}
b \ &=& \ L/\ell
\end{eqnarray}
where  $v_E=\sqrt{2 E / {\mathsf m}}$.
One difference between the two models is related
to the coupling parameter $\sigma$, which has been
defined in the beginning of Section~\ref{regimes1}
as the RMS value of the in-band off-diagonal elements:
\begin{eqnarray} \label{e22}
\sigma   \  =& \  (\ell/L) V_0  & \ \ \ \ \mbox{for the bump}
\\ \label{e23}
\sigma  \ =& \  (\ell/L)^{1/2} V_0  & \ \ \ \  \mbox{for the disorder}
\end{eqnarray}
This leads to
\begin{eqnarray} \label{e24}
\delta x_c^{\rm bump}  \ &=  \  \Delta/\sigma \ = \  (\Delta_b/V_0)
\\ \label{e25}
\delta x_c^{\rm disorder}  \ &=  \  \Delta/\sigma \ = \  (\Delta_b/V_0)  / \sqrt{b}
\end{eqnarray}
However, the difference in $\sigma$, and hence in $\delta x_c$
is not the significant difference between the ``bump"
potential and the ``disorder" potential. The significant
difference is related to the {\em statistical properties} of the
$\mbf{B}$ matrix. In the case of ``disorder" the
matrix elements look ``random", which is not
the case for a single bump.

In both cases, of having either single bump or disorder,
the FOPT regime is $\delta x \ll \delta x_c$.
What do we have beyond FOPT?
More specifically, the question is whether we have,
as in the theory of chaotic systems, a distinct parametric scale
$\delta x_{prt} =  \sqrt{b} \delta x_c$ that distinguishes
between a ``Wigner regime" and a ``semiclassical regime".
In the next section we shall try to address this question
within the framework of the uniform approximation.

\section{\label{uniform_regimes}Regimes within the framework of the uniform approximation}

There is a general semiclassical procedure that associates
wavefunctions of integrable systems with phase space manifolds.
The traditional implementation of this procedure, in case of
one dimensional systems, is known as ``the WKB method".
The WKB method is problematic near turning points.
This problem can be solved by ``uniformization" of the solution.
The simplest point of view regarding this ``uniformization"
is obtained by using action-angle variables. This leads
to a Hamiltonian that looks like that of a particle in a ring.
In general the dispersion relation can be different
(not quadratic as in (\ref{e5})), but we shall see that
this is not an important difference for the issues under study.

The eigenstates of Hamiltonian (\ref{e5}) are supported by the
manifolds $H(Q,P;x)=E_n$. An alternate (explicit) way
to describe a given manifold is $P=p_n(Q;x)$, where
\begin{eqnarray} \label{e26}
p_n(Q; x)  \ = \
\sqrt{2{\mathsf m}\left[ E_n{-}\delta xV(Q) \right]}
\ \approx \
p_n - \  \delta x \ \hbar \ k(Q)
\end{eqnarray}
and $k(Q)=V(Q)/\hbar v_E$. In the following we
use units such that $\hbar=1$.
The semiclassical formula for the corresponding eigenfunction is
\begin{eqnarray} \label{e27}
\langle Q|n(x)\rangle \ = \  \frac{1}{\sqrt{L}}
\exp\left(i \int_0^Q p_n(Q';x)dQ' \right)
\end{eqnarray}
This can be regarded as  a special case of (\ref{eC5}).
Above we approximate the classical pre-exponential prefactor
by  ${1}/{\sqrt{L}}$. This is legitimate because a fixed assumption of
this Paper is that we are considering high lying eigenstates,
and assume {\em classically small} perturbations 
\footnote{
It can also be regarded as an example of a more general semiclassical
perturbation approximation \cite{miller2,fdl2}.}.
We already saw, in the context of the deformable harmonic oscillator
(see the end of \ref{appC}), that the numerical error which is associated
with this approximation is insignificant. The essential physics of
having various ``regimes" is not related to this approximation.

The overlap of the semiclassical wavefunctions is given
by the integral
\begin{eqnarray} \label{e28}
\hspace*{-2cm}
\langle n(x) | m(x_0) \rangle = \frac{1}{L} \int_0^L dQ
\ \exp\left[- i\left( (p_n - p_m)Q - \delta x \int_0^Q k(Q')dQ' \right)\right]
\end{eqnarray}
In the following subsections we discuss the consequences 
of this expression in case of either bump or disorder.

\subsection{bump case} 

In case of the bump, $k(Q)$ has an amplitude
$k_0=V_0/\hbar v_E$ over a spatial scale $\ell$.
The total phase variation in (\ref{e28}) is
\begin{eqnarray} \label{e29}
\delta\phi_{\rm bump} \ = \   \delta x   \times   k_0 \ell
\end{eqnarray}
This phase variation is in fact the
phase space area of the bump.
So we have two possibilities:
Either $\delta\phi_{\rm bump} \ll 2\pi$
or $\delta\phi_{\rm bump} \gg 2\pi$.
This can be shown to be equivalent
to either $\delta x \ll \delta x_c^{\rm bump}$ or
$\delta x \gg \delta x_c^{\rm bump}$ respectively.
In the former case it is easy to recover
the FOPT result (\ref{e17}).
One should simply put the perturbation
off the exponent, and then do integration
by parts. For $n\ne m$ it leads to
\begin{eqnarray} \label{e30}
\hspace*{-1cm}
P(n|m) = \left|
\frac{1}{L}\int_0^L dQ
\frac{\eexp{- i (p_n - p_m)Q}   }{p_n-p_m}
\ \delta x k(Q)
\right|^2 \ = \
\delta x^2 \left|\frac{\mbf{B}_{nm}}{E_n-E_m} \right|^2
\end{eqnarray}
The other possibility ($\delta\phi_{\rm bump} \gg 2\pi$)
guarantees the validity of the standard
semiclassical approximation. In such case
we cannot put the perturbation off the exponent,
but instead we can make a stationary phase
approximation. We shall not dwell further
on details because it is a standard textbook procedure.

\subsection{disorder case}

We see that for a simple bump we have either
FOPT or semiclassical approximation.
So we have just two regimes, in spite of the fact
that $b\gg 1$.
This means that large bandwidth is not a sufficient
condition for having three parametric regimes.
So let us try to make things more complicated by
considering a disordered potential with many bumps.
Equation~(\ref{e29}) still describes the phase variation over a single bump.
This means that $\delta\phi_{\rm bump} \gg 2\pi$ is still
the relevant condition for a semiclassical approximation!
What about first order perturbation theory?
The total phase variation in  (\ref{e28})  for many bumps is
\begin{eqnarray} \label{e31}
\delta\phi_{\rm disorder} =   \delta x \times \sqrt{b} \times   k_0 \ell
\end{eqnarray}
(note that $b$ is essentially the number of bumps involved).
Consequently the validity condition for first order perturbation theory
is $\delta\phi_{\rm disorder} \ll 2\pi$, which can be easily converted
into $\delta x \ll \delta x_c^{\rm disorder}$.

The considerations of the previous paragraph imply
that for {\em disordered} potential we have three regimes.
In the intermediate regime
$\delta x_c \ll \delta x \ll \delta x_{\tbox{prt}}$,
we have $\delta \phi_c^{\rm disorder} \gg 2\pi$
while $\delta \phi_c^{\rm bump} \ll 2\pi$.
Consequently we can use neither FOPT, nor the semiclassical approximation.
This is the Wigner regime where we expect to find a  Lorentzian-like LDOS.
Let us demonstrate that indeed a  Lorentzian-like LDOS
can be obtained from the uniform approximation (\ref{e28}).
For this purpose we average $P(n|m)$ over realizations of the disorder:
\begin{eqnarray}
\hspace*{-2.5cm}
P(n|m) = \frac{1}{L^2}\int_0^L\int_0^L dQ_1dQ_2
\ \eexp{-i(p_n-p_m)(Q_2-Q_1)}
\left\langle \exp\left[i\delta x\int_{Q_1}^{Q_2}k(Q')dQ'\right] \right\rangle_{\rm disorder}
\nonumber \\ \hspace*{-1.3cm}
= \frac{1}{L^2}\int_0^L\int_0^L dQ_1dQ_2
\ \eexp{-i(p_n-p_m)(Q_2-Q_1)}
\exp\left[-\frac{1}{2}  \left(\frac{\delta x }{\hbar v_E}\right)^2
\int_{Q_1}^{Q_2} \int_{Q_1}^{Q_2}\langle V(Q')V(Q'')\rangle dQ'dQ'' \right]
\nonumber \\ \hspace*{-1.3cm}
\approx \frac{1}{L}\int_{-\infty}^{\infty} dr \ \eexp{-i(p_n-p_m)r}
\exp\left[-\delta x^2  \left(\frac{V_0}{\hbar v_E}\right)^2 \ell |r| \right]
\label{e32}
\end{eqnarray}
The last integral is the Fourier transform of an exponential,
leading to the Lorentzian LDOS as defined in (\ref{e18}).

\section{\label{strong_localization}Strong localization effect}

Still we have to address the question whether we can trust
the uniform approximation, which is based on the WKB wavefunction (\ref{e27}).
The answer is known to be negative in case of {\em disordered}
potential. The WKB approximation does not take into account
backscattering, which is responsible for the strong localization
effect in 1D disordered potential. It is well known \cite{richter} that
the localization length is equal (in 1D) to twice the mean
free path. Up to a numerical prefactor the Born approximation estimate is
\begin{eqnarray} \label{e33}
\hspace*{-1cm}
\frac{1}{L_{\tbox{loc}}}
\ &\approx & \
\frac{1}{\hbar v_E} \frac{\sigma^2}{\Delta}
\times \delta x^2
\ = \  \left( \frac{\delta x}{\delta x_c}\right)^2 \frac{1}{L}
\ = \  \left( \frac{\delta x}{\delta x_{\tbox{prt}}}\right)^2 \frac{1}{\ell}
\end{eqnarray}
The condition for {\em not} being affected by the strong localization
effect is $L_{\tbox{loc}} \gg L$, which  leads to $\delta x \ll \delta x_c$.
Thus it follows that only in the FOPT regime we can ignore
the strong localization effect. The strong localization  effect cannot be
ignored neither in the Wigner regime nor in the semiclassical regime.

The above Born approximation assumes $L_{\tbox{loc}} \gg \ell$.
This condition breaks down if $\delta x > \delta x_{\tbox{prt}}$.
Recall that we also assume that $\delta x$ is small in the
classical sense ($\delta x V_0 \ll E$). The two inequalities are consistent
if and only if the de Broglie wavelength of the particle is much smaller
compared  with $\ell$. In this regime we can analyze the localization
using the well known transfer matrix approach: Each bump has
some transfer matrix, and the random distance between the bumps
provides the phase randomization which is assumed in ``combining"
adjacent transfer matrices. Denoting the
average transmission of a ``bump"  by $g$, one gets
\begin{eqnarray} \label{e33a}
\frac{1}{L_{\tbox{loc}}} \ \approx \ \ln(1/g) \times \frac{1}{\ell}
\end{eqnarray}
Thus even in the  $\delta x > \delta x_{\tbox{prt}}$ regime
we can have a very long localization length ($L_{\tbox{loc}} \gg \ell$),
which is in fact consistent with the naive expectation.

The implications of the above discussion are, that in spite of
the strong localization effect, it is still meaningful to distinguish
between three parametric regimes (FOPT, Wigner, semiclassical).
We just have to remember that the wavefunction is not ergodic
in real space, so the role of $L$ is taken by  $L_{\tbox{loc}}$.

\section{\label{survival}The survival probability ${\cal P}(t)$ }

We turn to discuss the calculation
of the survival probability (\ref{e1})
for the specific ``wavepacket dynamics" scenario
that has been defined in Section~\ref{definitions}.
We have the following five strategies of calculation: \\
\begin{minipage}{\hsize}
\vspace*{0.1cm}
\begin{itemize}
\setlength{\itemsep}{0cm}
\item
Uniform approximation  (which is essentially exact)
\item
Time domain classical  approximation
\item
Energy domain classical approximation  ($\leadsto$ LDOS $\leadsto$  Fourier transform)
\item
Time domain perturbation theory
\item
Energy domain perturbation theory ($\leadsto$ LDOS $\leadsto$  Fourier transform)
\end{itemize}
\vspace*{0.0cm}
\end{minipage}

It can be shown that in typical circumstances
the two versions of perturbation theory give in leading order consistent results.
This means that we can write a perturbative (essentially first order) result that can be trusted
for sufficiently short times (for any perturbation~$\delta x$)
or for sufficiently weak perturbation (for any time~$t$).
As an example we consider the deformable harmonic oscillator.
Taking equation~(\ref{e15}) with appropriate second order
compensation of normalization, we get after Fourier transform,
\begin{eqnarray} \label{e34}
{\cal P}^{\tbox{prt}}(t) \ \approx \ 1 - 2\left(\frac{\delta x}{x}\right)^2 E^2  \sin^2 t
\end{eqnarray}
In a strict time domain FOPT we get only $t^2$ time
dependence, while in a strict energy domain FOPT
we do not get the correct normalization (${\cal P}^{\tbox{prt}}(0) =1$).
Still we are able to get one consistent result.
The situation is different with the classical
approximation. Here time domain and energy domain
calculations do not give the same result.
As an example we consider again the deformable
harmonic oscillator. The calculation of overlap
between $\rho^{cl}_t$ and  $\rho^{cl}_0$ is
simpler in action-angle variables, but otherwise
it is similar to the calculation in \ref{appA}, leading to
\begin{eqnarray} \label{e35}
{\cal P}^{cl}(t) \ = \  \left|2\pi\frac{\delta x}{x}E\sin t \right|^{-1}
\end{eqnarray}
The energy domain classical approximation is
obtained by squaring the Fourier transform of
equation~(\ref{e10}) using the classical approximation (\ref{e13}) of $P^{cl}(n|m)$, leading to
\begin{eqnarray} \label{e36}
{\cal P}^{cl,E}(t) \ = \ \left|J_0\left(2\frac{\delta x}{x} Et\right)\right|^2
\end{eqnarray}
Although there is no simple exact expression for ${\cal P}(t)$,
there again exists a very simple uniform approximation which is remarkably accurate
in all regimes and therefore we can regard it as  ``exact,'' namely
\begin{eqnarray} \label{e37}
{\cal P^{\rm uniform}}(t) \ = \ \left|J_0\left(2\frac{\delta x}{x} E \sin t \right)\right|^2
\end{eqnarray}
It is derived in \ref{appD} by using semiclassical expressions for the initial and evolved states,
while calculating the overlap exactly rather than by the stationary phase approximation.

\begin{figure}
\centerline{\epsfig{figure=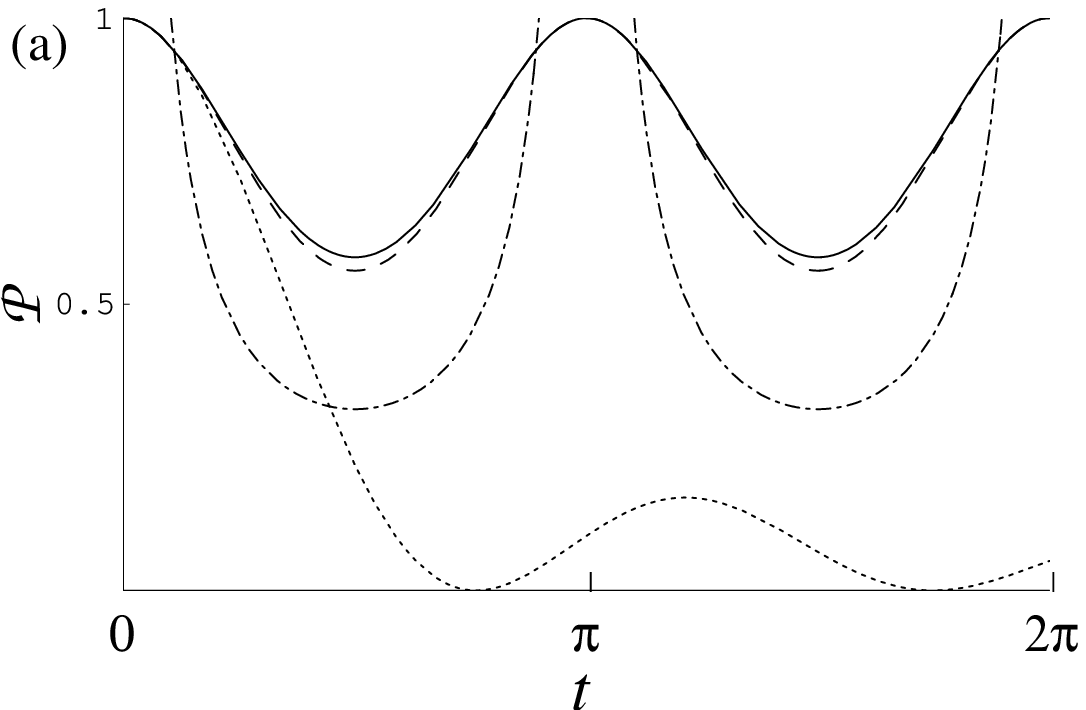,width=0.45\hsize}} \ \\
\centerline{\epsfig{figure=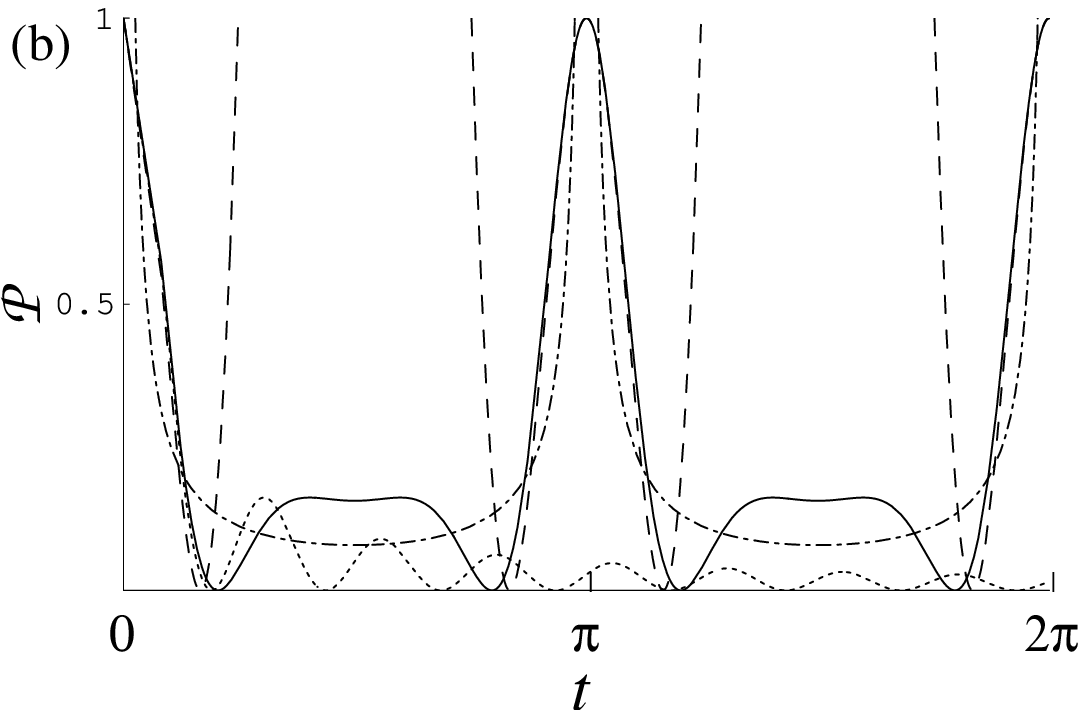,width=0.45\hsize}} \ \\
\centerline{\epsfig{figure=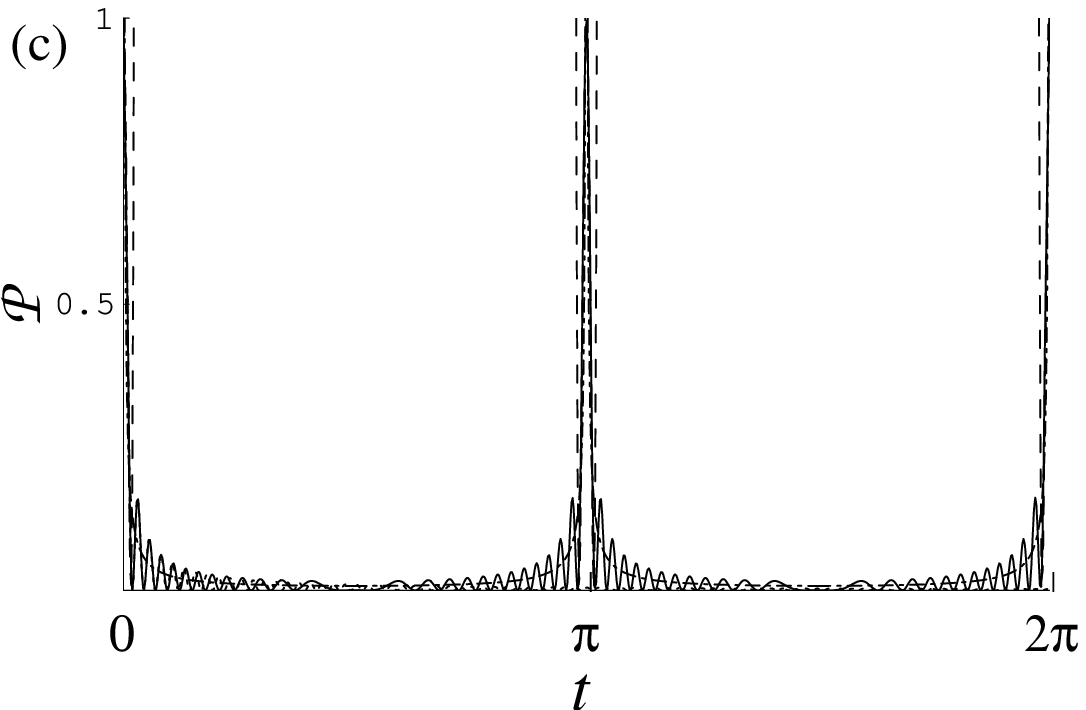,width=0.45\hsize}} \ \\
\centerline{\epsfig{figure=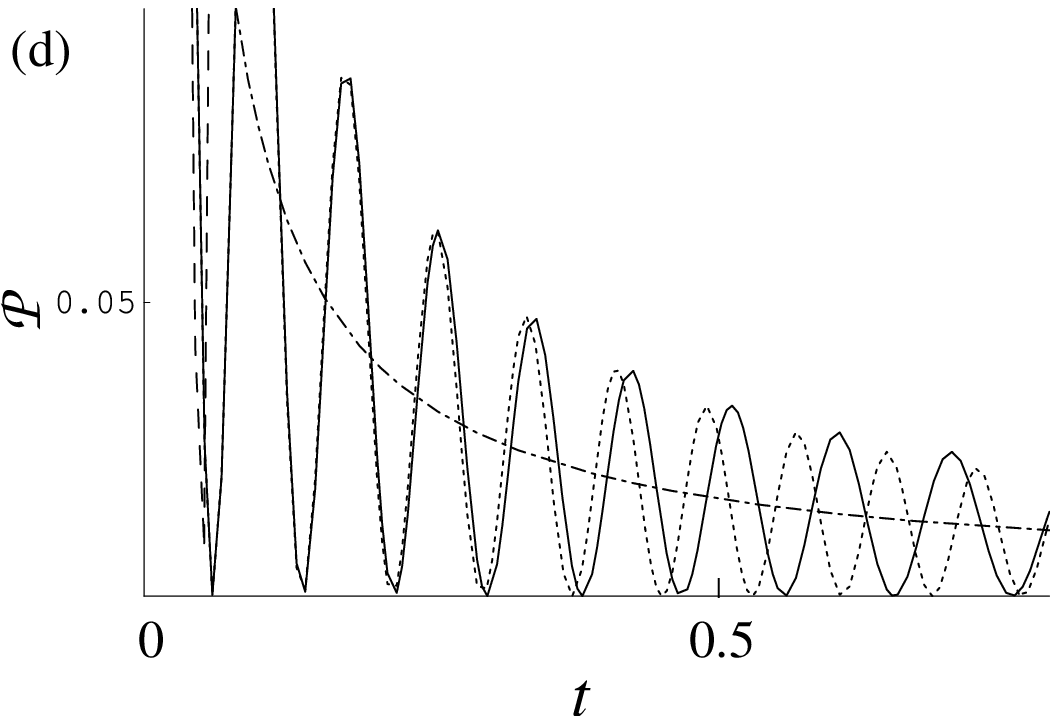,width=0.45\hsize}} \ \\

\caption{\label{fig_survival}
The survival probability, using the same parameters as in figure~\ref{fig_ldos}.
(a) Perturbative regime.
(b) Intermediate regime.
(c) Semiclassical regime.
In each panel the solid line is the uniform result (\ref{e37}),
the dashed line is the perturbative result (\ref{e34}),
the dashed-dotted line is the time domain classical result (\ref{e35}),
and the dotted line is the energy domain classical result (\ref{e36}).
Panel (d) gives a short-time detail of panel (c).
}

\end{figure}

Results (\ref{e34})-(\ref{e37}) are graphically displayed in figure~\ref{fig_survival}. 
Approximations (\ref{e34})-(\ref{e36}) can be regarded as various limits of the uniform approximation (\ref{e37}).
It is easily seen that (\ref{e37}) reduces to the perturbative result (\ref{e34})
whenever $2 (\delta x / x ) E\sin(t) \ll 1$.
So as expected the perturbative result can be trusted for either small time $t$
or for small perturbation $\delta x / x$.
It is also easy to see that the uniform result (\ref{e37}) reduces
to the energy domain classical result (\ref{e36}) for $t \ll 1$.
The relation of (\ref{e37}) to the time domain classical result (\ref{e35})
is more subtle: For large perturbation the two expressions agree in an asymptotic sense,
and either time smoothing  or energy averaging is required in order
to demonstrate this agreement (see figure~\ref{fig_survival}c).

An important message of this section is that the discussion of regimes
in the context of ${\cal P}(t)$ is in one to one correspondence
with the discussion of LDOS regimes.

\section{\label{conclusions}Conclusions and final remarks}

In general, for more complicated ``wavepacket dynamics" scenarios, 
which may involve time dependent Hamiltonian,  
an explicit reduction of the survival probability problem  
to LDOS study is not possible \cite{fdl}. 
The simplest way to make the Hamiltonian ``time dependent" 
is by changing it either once (as in ``fidelity" or ``Loschmidt echo" studies) 
or repeatedly (as in kicked systems).  
The latter possibility opens the way for ``chaos" 
in the classical dynamics of one-dimensional (1D) systems.
A prototype system for 1D chaos studies is the kicked rotator:
This is a particle in a ring, which is periodically kicked by a cosine potential.
As in the standard formulation of ``wavepacket dynamics" it is common
to assume an initial preparation which is supported by a manifold $P=p_m$.
As a result of the kicks the initial manifold becomes extremely convoluted.
If we want to calculate ${\cal P}(t)$ we have to trace
the evolving manifold with the initial one. The calculation
of this overlap can be carried out using the uniform approximation
that we have discussed previously~\cite{jiri,jiri2,fdl2}.

There is one major difference that makes the ``chaotic" scenario
that we have described above different from the LDOS calculation:
The issue of strong localization, which has been
discussed in Section~\ref{strong_localization}  is no longer relevant.
The evolving state is ``distributed" over the whole manifold
\footnote{
The only reservation for this statement is the possibility
of witnessing a ``dynamical localization effect" after an extremely
long time (known as the ``breaktime").
For small $\hbar$ the existence of a ``breaktime" will have
a negligible effect on the behavior of ${\cal P}(t)$.
In fact we should remember that also in non-kicked systems
we have a ``breaktime", which is just the Heisenberg time.
We also note that this type of localization is apparently
captured by semiclassical methods \cite{kaplan}}.

{\em In view of the above, the calculations that we have presented
in this paper, using the uniform approximation, are in fact generic}.
In particular we have explained how a Fermi-golden-rule behavior
arises within this framework. A refinement
of the uniform approximation using the
``replacement manifolds" approach  has been introduced
in references \cite{jiri,jiri2}.

The reader can be tempted to reach the conclusion that
the discussion of ``regimes" within the framework
of the ``uniform approximation" can be trivially generalized 
from $d=1$ chaotic systems to $d>1$ chaotic systems. 
This is in fact not quite correct.
In order to explain the difficulty let us consider
again LDOS calculation. In one-dimensional systems
what we have to do is to calculate the overlap of
two one-dimensional manifolds.  We can do this
calculation semiclassically if we can trust the
stationary phase approximation. This leads to the
"$\hbar$ area" condition: the stationary phase approximation
is accurate if the phase space area delimited by the two manifolds between
two stationary phase points is larger than $\hbar$.
What is the generalization of
this condition in the $d>1$ case? Now the manifolds
are ``surfaces" and they intersect along ``lines".
So the concept of ``stationary points" becomes
inapplicable, and also the ``$\hbar$ area" condition
becomes meaningless. The way out of this difficulty \cite{vrn}
is to use the Wigner function point of view. Then one realizes
that the proper way to formulate the semiclassical condition
is to say that the separation between the surfaces should be much
larger compared with the ``thickness" of Wigner function.
This ``thickness" is just the ``bandwidth" that we have
discussed in this Paper.

Conventionally, semiclassical methods are applied
in the ``semiclassical" regime (large enough perturbation).
In the perturbative regime people use perturbation theory.
A major motivation  for the present line of study is
to extend the applicability of semiclassical methods into
the perturbative regime. Most of the calculations that we have
presented under the heading ``uniform approximation" can be
reformulated using the Wigner function language.
This opens the way towards a unified semiclassical
understanding of ``regimes" in case of  $d>1$  systems.

\ack

It is our pleasure to thank Bilha Segev (BGU) and Eric Heller (Harvard)
for useful discussions. The research was supported by
the Institute for Theoretical Atomic and Molecular Physics at Harvard University,
by the Mathematical Sciences Research Institute at Berkeley, and
by the Israel Science Foundation (grant No.11/02).

\appendix

\section{\label{appA}Calculation of $P^{cl}(n|m)$ and $P^{\rm exact}(n|m)$}

We shall demonstrate in this appendix the calculation of $P^{cl}(n|m)$
for the deformable harmonic oscillator:
\begin{eqnarray} \label{eA1}
P^{cl}(n|m) &
\ = \  \int\frac{dQdP}{2\pi} \delta\left[H(Q,P;x_2)-E_n\right]\delta\left[H(Q,P;x_1)-E_m\right]
\nonumber \\ &
\ = \ \int\frac{dQ}{2\pi} \sum_{P = \pm \sqrt{2 E_{m} - (Q / x_1)^2}/ x_1}
\frac{\delta [ {\cal H}(Q, P; x) - E_n ]}{x_1^2 |P|}
\nonumber \\ &
\ = \  \frac{1}{\pi} \int dQ
\frac{\delta [f(Q)]}{\sqrt{2 x_1^2E_m - Q^2}}
\nonumber \\ &
\ = \ \frac{1}{\pi} \sum_{\pm} \left ( 2 x_1^2 E_m - Q_{\pm}^{2} \right)^{-1/2} \left| f'(Q_{\pm}) \right|^{-1}
\nonumber \\ &
\ = \ \frac{1}{\pi} \frac{x_1 x_2}
{\sqrt{\left(x_{2}^{2} E_n - x_{1}^{2} E_m \right) \left(x_{2}^{2} E_m - x_{1}^{2} E_n \right)}}
\end{eqnarray}
where
\begin{eqnarray} \label{eA2}
f(Q) \ \equiv \ \frac{x_{2}^{4} - x_{1}^{4}}{2 x_{1}^{4} x_{2}^{2}} Q^2 + \left
( \frac{x_2}{x_1} \right)^2 E_m - E_n.
\end{eqnarray}
and $Q_{\pm}$ are the roots of the equation $f(Q)=0$,
\begin{eqnarray} \label{eA3}
Q_{\pm} \ = \ \pm x_{1} x_{2} \sqrt{\frac{2 (x_{2}^{2} E_m - x_{1}^{2} E_n)}{x_{2}^{4} - x_{1}^{4}}}
\end{eqnarray}
for which
\begin{eqnarray}
f'(Q_{\pm}) \ = \ \pm x_1^{-3} x_2^{-1} \sqrt{2 (x_{2}^{2} E_m - x_{1}^{2} E_n) (x_{2}^{4} - x_{1}^{4})}
\end{eqnarray}
Equation~(\ref{e13})  is a simplified version of this result,  assuming that $\delta x \ll x$.

It is also possible to obtain an exact result in the quantum mechanical case.
The explicit expression for the eigenfunction using Hermite polynomials is
\begin{eqnarray}
\bra Q | n(x) \ket \ = \ (\pi x^2)^{-1/4} (2^n n!)^{-1/2} H_n(Q/x)
\eexp{-(Q/x)^2 / 2}
\end{eqnarray}
This leads to
\begin{eqnarray} \nonumber
& \langle n(x_2)|m(x_1) \rangle =  (\pi x_1 x_2)^{-1/2}
(2^{n+m}n!m!)^{-1/2} \nonumber
\\ \label{eA4}
&  \ \ \times   \int_{-\infty}^{\infty}    dQ H_n(Q / x_2) H_m(Q / x_1)
\eexp{ -\half (x_1^{-2} + x_2^{-2} ) Q^2}
\end{eqnarray}
Upon squaring one obtains $P(n|m)$. The integral in (\ref{eA4}) becomes highly oscillatory
for high-lying eigenstates in which we are interested, and numerical calculation is tricky.
To our surprise, this intimidating integral can be evaluated analytically even for $x_1 \neq x_2$,
resulting in a finite sum of terms, which is not very elegant, but very simple to evaluate on a computer.
For the exact benchmark in the numerical results presented in this paper,
we therefore used this analytical expression instead of numerically evaluating the integral (\ref{eA4}).

Note that both classically and quantum mechanically
the overlap $P(n|m)$ depends only on the ratio $x_2/x_1$.
In the classical case $n$ is a real index ($E_n$ can have
any real positive value). In the quantum mechanical
case $n$ is an integer index ($E_n= n + \half$).
Due to the reflection symmetry of the Hamiltonian
there is an overlap only between states with the same
parity. The overlap for $|n-m|=\mbox{odd}$  vanishes.
Whenever we say that $P(n|m) \approx P^{cl}(n|m)$,
it should be interpreted in a coarse grain sense.
Therefore, for sake of graphical presentation we have
plotted $P(n|m)$ versus $P^{cl}(n|m)/2$.
One may  say that the plotted $P(n|m)$ and $P^{cl}(n|m)/2$
correspond to a de-symmetrized oscillator.

\section{\label{appB}Calculation of $P^{\tbox{prt}}(n|m)$}

We shall demonstrate in this appendix the calculation of $P^{prt}(n|m)$
for the deformable harmonic oscillator. From first order perturbation theory
we know that for $m \ne n$
\begin{eqnarray}
\langle m(x_0) | n(x_0+\delta x) \rangle \approx
\frac{ \langle m |\delta {\cal H} | n \rangle }{E_n-E_m }
\end{eqnarray}
Using
\begin{eqnarray} \nonumber
\hspace*{-2.5cm}
& \bra m \vert  Q^2  \vert n \ket   ,\    \bra m \vert P^2 \vert n \ket
\ = \  (\half + n) \delta_{mn}
& \pm \frac{1}{2}
\left[\sqrt{(n+1)(n+2)} \delta_{m, n+2} + \sqrt{n (n-1)} \delta_{m,n-2} \right]
\end{eqnarray}
one obtains
\begin{eqnarray}
P^{\tbox{prt}}(n|m) \ \approx \
\delta_{n,m} + \frac{1}{4}\left(\frac{\delta x}{x}\right)^2 \left(E^2-\frac{1}{4}\right) \delta_{|n-m|,2}
\end{eqnarray}
where $E=(E_n+E_m)/2$. In the text we have presented
a simplified version that assumes $E \gg 1$.

\section{\label{appC}Calculation of $P^{\rm uniform}(n|m)$}

The following canonical transformation is used in order
to transform the Hamiltonian of the harmonic oscillator
to action-angle variables:
\begin{eqnarray}
Q \ &=& \ \sqrt{2I} \cos(\phi) \\
P \ &=& \ \sqrt{2I} \sin(\phi)
\end{eqnarray}
In the vicinity of $x_0=1$ it leads to
\begin{eqnarray} \nonumber
{\cal H}(\phi,I; x)  \ &=& \
I \left[(1/x)\cos^2\phi+x\sin^2\phi\right]
\\ \label{eC3}
\ & \approx & \
I - \delta x I \cos(2\phi)) + {\cal O}(\delta x^2)
\end{eqnarray}
The canonical transformation from $(\phi,I)$ to the
action angle variables $(\phi',I')$ of the perturbed
Hamiltonian is derived from a generating function $S(\phi,I')$.
The manifold $I'=\const$ is determined from
the equation ${\cal H}(\phi,I; x)=\const$,
leading to the relation $I = \left[ 1+\delta x \cos(2\phi) \right] I'$.
The generating function should satisfy
$I={\partial S}/{\partial \phi}$. Therefore one deduces that
\begin{eqnarray}
S(\phi,I') \ = \ \left[ \phi+\delta x \sin(2\phi) \right] I'
\end{eqnarray}
The semiclassical expression for the wavefunction is
\begin{eqnarray} \label{eC5}
\bra \phi \vert I' \ket_{\rm sc} \ = \  (2 \pi)^{-1/2} \sqrt{\frac{\partial^2 S}{\partial I' \partial \phi}} \ \eexp{iS(\phi,I')}.
\end{eqnarray}
The $n$th semiclassical eigenstate corresponds
to the substitution $I'=E_n=n + \half$.
For technical simplicity it is convenient to
calculate the overlap between two perturbed
wavefunctions ($x=x_0 \pm \delta x/2$) .
\begin{eqnarray} \nonumber
\hspace*{-2cm}
\langle n(x_0 + \delta x / 2 ) | m(x_0 - \delta x / 2 ) \rangle
\  = \ \int_0^{2\pi} d\phi \langle \phi | I''{=}E_n \rangle^{*} \langle \phi | I'{=}E_m \rangle
\\ \nonumber
 = \frac{1}{2\pi} \int_0^{2\pi} d\phi  \sqrt{1 - \delta x^2 \cos^2 2\phi}
\times \exp \left[ i (E_m - E_n) \phi - i  \delta x E  \sin 2 \phi \right]
\\
\approx J_{(m-n)/2}(\delta x E)
\end{eqnarray}
Above $I'$ and $I''$ correspond to the perturbations $\pm \delta x/2$,
and $E=(E_n+E_m)/2$. In the main text we have reverted to
a more general version of this expression that does not assume $x_0=1$.
It is important to realize that because of the assumption $\delta x \ll x$
the pre-exponential factor (which is in fact a ``classical" factor)
can be neglected in leading order.  Figure~\ref{fig_ldos} confirms the remarkable
agreement of the uniform approximation with the exact result.

\section{\label{appD}Calculation of ${\cal P}^{\rm uniform}(t)$}

For the purpose of calculation it is more convenient
to regard the prepared state as a ``perturbed state"
and the evolution Hamiltonian  as the ``unperturbed
Hamiltonian". This is of course equivalent to the presentation
in the text upon the replacement $\delta x\mapsto -\delta x$.

The initial state $|m(x_0+\delta x)\rangle$
is characterized by the action $I'=E_m\equiv E$.
It is represented as in \ref{appC} by
\begin{eqnarray}
\langle \phi | I'{=}E \rangle \approx \frac{1}{2\pi} \exp[i(\phi+\delta x\sin 2\phi) E]
\end{eqnarray}
The evolving state is represented by
\begin{eqnarray}
\langle \phi |\eexp{-it{\cal H}_0} | I'{=}E \rangle = \langle \phi-t | I'{=}E \rangle
\end{eqnarray}
The overlap between the evolving and the initial state is
\begin{eqnarray} \nonumber
\frac{1}{2\pi} \int_0^{2\pi} d\phi \exp \Big[ i \Big( -t + \delta x \left[ \sin 2 (\phi-t) - \sin 2 \phi \right] \Big) E \Big]
\end{eqnarray}
For $F(t)$ as defined in (\ref{e12}) we obtain
(after the required replacement $\delta x\mapsto -\delta x$):
\begin{eqnarray}
\hspace*{-2cm}
F(t) \  = \  \frac{1}{2\pi} \int_0^{2\pi} d\phi \exp \left[ i  \delta x E \sin t \cos(2 \phi-t) \right]
\  = \ J_0\left(2 \delta x E \sin t\right)
\end{eqnarray}
The survival probability is obtained by squaring this result.




\Bibliography{99}

\bibitem{wls}
Cohen D and Heller E J 2000 {\it Phys. Rev. Lett.} {\bf 84} 2841

\bibitem{heller}
Heller E J 1991 {\it Chaos and Quantum Systems},
ed M-J~Giannoni {\it et al} (Amsterdam: Elsevier)

\bibitem{bilha}
Kallush S, Segev B, Sergeev A V and Heller E J 2002
{\it J. Phys. Chem. A} {\bf 106} 6006

\par\item[] Sergeev A V and  Segev B 2002
{\it J. Phys. A: Math. Gen.} {\bf 35}  1769

\bibitem{fidelity}
Jalabert R A and Pastawski H M 2001
{\it Phys. Rev. Lett.} {\bf 86} 2490
%
\par\item[] Cucchietti F M, Pastawski H M, Jalabert R 2000
{\it Physica A} {\bf 283} 285
%
\par\item[] J.V. Emerson, Phys. Rev. Lett. {\bf 89}, 284102 (2002).
%
\par\item[] Cucchietti F M, Pastawski H M and Wisniacki D A 2002
{\it Phys. Rev. E} {\bf 65} 045206
%
\par\item[] Cucchietti F M, Lewenkopf H, Mucciolo E R,
Pastawski H M and Vallejos R O 2002
{\it Phys. Rev. E} {\bf 65} 046209
%
\par\item[] Benenti G and Casati G 2002 {\it Phys. Rev. E} {\bf 65} 066205
%
\par\item[] Cerruti N R and Tomsovic S
{\it Phys. Rev. Lett.} {\bf 88} 054103
%
\par\item[] Prosen T 2001 {\it Preprint} quant-ph/0106149
%
\par\item[] Prosen T and Znidaric M 2001 {\it J. Phys. A} {\bf 34} L681
%
\par\item[] Eckhardt B 2003 {\it J.Phys. A} {\bf 36} 371
%
\par\item[] Silvestrov P G, Tworzydlo J and Beenakker C W J 2002 {\it Preprint} nlin.CD/0207002

\bibitem{jacq}
Jacquod P, Silvestrov P G and Beenakker C W J 2001
{\it Phys. Rev. E} {\bf 64} 055203

\bibitem{fdl}
Wisniacki D A and Cohen D 2002 {\it Phys. Rev. E} {\bf 66} 046209

\bibitem{fdl2}
Van\'{\i}\v{c}ek J and Heller E J 2003 {\it Preprint} quant-ph/0302192



\bibitem{ovrv}
For a pedagogical presentation, including references,
see references \cite{dsp} and \cite{vrn}, which can be
downloaded from http://www.bgu.ac.il/$\sim$dcohen.


\bibitem{dsp}
Cohen D 2002
"Driven chaotic mesoscopic systems, dissipation and decoherence"
in
{\it Dynamics of Dissipation: Proceedings of the 38th Karpacz Winter School of
Theoretical Physics}
ed P Garbaczewski and R Olkiewicz (Springer)

\bibitem{vrn}
Cohen D 2000
"Chaos, dissipation and quantal Brownian motion" in
{\it New Directions in Quantum Chaos: Proceedings of the International School of Physics Enrico Fermi Course CXLIII}
ed G Casati, I Guarneri and U Smilansky (Amsterdam: IOS Press)

\bibitem{crs}
Cohen D 1999 {\it Phys. Rev. Lett.} {\bf 82} 4951

\bibitem{jiri}
Van\'{\i}\v{c}ek J and Heller E J 2001 {\it Phys. Rev. E} {\bf 64} 026215

\bibitem{jiri2}
Van\'{\i}\v{c}ek J and Heller E J 2003 {\it Phys. Rev. E} {\bf 67} 016211

\bibitem{lds}
Cohen D and Kottos T 2001 {\it Phys. Rev. E} {\bf 63} 36203

\bibitem{miller} 
Miller W H 1974 {\it Adv. Chem. Phys.} {\bf 25} 69

\bibitem{miller2} 
Miller W H and Smith F T 1978 {\it Phys. Rev. A} {\bf 17} 939

\bibitem{richter}
Richter K, Ullmo D and Jalabert R A 1996 {\it Phys. Rev. B} {\bf 54} R5219

\bibitem{kaplan}
L. Kaplan, Phys. Rev. Lett. {\bf 81}, 3371 (1998).

\end{thebibliography}
\end{document}